# Metal-Dielectric Bilayer Supported Nanoporous Silver Substrates for Ultrasensitive Surface-Enhanced Raman Scattering

*Maryam Naseer[1,2]‡, Anastasiia Sapunova[1]‡, Ali Douaki[1], Vincenzo Caligiuri[3], Silvia Dante[4], Denis Garoli[1], and Roman Krahne[1]**

[1]Optoelectronics, Istituto Italiano di Tecnologia, Genova, Italy

[2]Dipartimento di Chimica e Chimica Industriale (DCCI), Università di Genova, Genova, Italy

[3]Dipartimento di Fisica, Università della Calabria, Cosenza, Italy

[4]Materials Characterization Facility, Istituto Italiano di Tecnologia, Genova, Italy

ABSTRACT Resolving the chemical footprint of molecules in analytes is of fundamental interest in biosensing and biomedical applications. Surface-enhanced Raman Scattering (SERS) spectroscopy is a powerful tool for non-invasive molecule investigation, and there is a continuous effort to develop highly efficient SERS substrates that can be fabricated at low cost and tuned to the desired spectral range of investigation. Here, we show that the Raman signal of porous silver films can be significantly enhanced by integrating a thin dielectric layer and a metallic backreflector underneath the porous metal film. We call this architecture nanoporous metal on dielectric metal (NPMoDM) and show that it leads to much stronger near-field enhancement in the bulk of the film. Additionally, the spectral band of enhanced sensitivity can be tuned by the thickness of the dielectric layer within the visible spectral range. With rhodamine 6G and para-aminothiophenol as target analytes, we demonstrate SERS detection of substrates incubated with zeptomolar concentrations of the analyte, which corresponds to only few molecules present on a mm$^2$ surface area. Our approach introduces a tunable and scalable (lithography-free) platform for ultrasensitive SERS applications, with a readily accessible surface for analyte functionalization.

## Introduction

Surface enhanced Raman spectroscopy (SERS) is one of the key non-invasive tools in molecular biosensing due to its capability to measure the fingerprint of analytes. [1-3] To enable efficient SERS investigations, nanostructured metal films and surfaces can be designed that feature strong electromagnetic near-field enhancement.[4, 5] One popular approach towards nanostructured morphologies has been the deposition and/or assembly of colloidal plasmonic nanoparticles. Silver nanostructures are particularly advantageous for strong plasmonic field enhancement in the visible, and single molecule sensitivity has been achieved using silver colloids for the detection of rhodamine 6G (R6G) [6], and crystal violet.[7] Another approach consists of the top-down fabrication of nanostructures to achieve extreme near-field intensities in nanogaps. Arrays of metal nanostructures such as nanostar[8] and microflower[9] geometries generate multiple efficient hot spots and enable to detect Raman signal from zeptomolar concentrations of analytes chemisorbed on the substrates.[10] To confirm single molecule detection capability, the bi-analyte method,[11] and distinguishing of R6G isotopologues [12] have been reported. Mostly, such studies profited from the resonant behavior of the analytes with respect to the laser excitation, but also single molecule detection of non-resonant molecules such as adenine has been achieved.[13]

Nanoporous metal films have attracted strong interest towards large scale and low-cost approaches for efficient electromagnetic field enhancement, since typically they can be fabricated in few process steps and without the need of expensive and time-consuming lithography processes.[14-22] Recently, our group extended the fabrication of nanoporous silver films obtained a by "dry etching" method first proposed by Kwon et al. [23]. The method relies on plasma etching of a sacrificial polymer layer and enables the fabrication of porous layers of different metals,[24] and also of multilayer structures.[25] Inspired by metal-dielectric-metal nanocavities,[26] and by the strong field and signal enhancing impact of underlying dielectric-metal bilayers combined with metallic hemispheres[27] and Au nanohole arrays for chiral biosensing [28], we develop in this work nanoporous metal on dielectric-metal (NPMoDM) substrates that combine the spectral tunability and field enhancement of metal-dielectric-metal nanocavities with the accessible and strongly nearfield enhancing features of nanoporous metal films. Finite element methods simulations show that the integration of the dielectric-metal bilayer leads to full extension of the strong near-field enhancement into the volume of the porous metal film. Furthermore, the calculations show that the spectral behavior of the hotspots does not coincide with the minima in reflectance, and evidence the occurrence of hybrid plasmonic and cavity modes. We use R6G and para-aminothiophenol (p-ATP) as analytes to assess the performance of the NPMoDM substrates in SERS. The thickness of the dielectric layer determines the cavity resonances of the substrates (which appear as minima in reflectance), and by fabricating cavities with resonances centered at 510 nm and 640 nm we tailor the spectral band of field enhancement to the laser excitation lines at 514 nm and 633 nm of our Raman spectroscopy setup. We obtain reliable signal that clearly features the characteristic Raman peaks of R6G and p-ATP at analyte concentrations for which only few molecules can be expected on $mm^2$ surface area, confirming the robust performance of the NPMoDM substrates in ultrasensitive Raman spectroscopy.

## Results and Discussion

**Figure 1a** illustrates the architecture of the NPMoDM SERS substrates consisting of a porous Ag film deposited on a dielectric-metal bilayer. **Figure 1b** shows the reflectance recorded from the a bare nanoporous Ag film, and of NPMoDM substrates with two different dielectric layer thicknesses (Cavity 1 and Cavity 2). The porous Ag morphology sustains plasmonic near-field enhancement around 400 nm wavelength for the bare porous Ag film, as demonstrated by finite element methods simulations in **Figure S1**. With addition of the dielectric-metal bilayer, the NPMoDM develops two minima in reflectance, a narrow high frequency and a broad low-frequency one. [27] The center position of the broad low-frequency minimum can be tuned by the thickness of the dielectric layer, which we demonstrate by fabricating two NPMoMD structures with 60 nm and 80 nm dielectric layer thickness, called Cavity 1 and Cavity 2, respectively. The broad resonances of Cavity 1 and Cavity 2 are tailored to the laser excitation wavelengths of 514 nm and 633 nm of our Raman spectroscopy setup. Furthermore, also the narrow high frequency reflectance minima show a minor dependence on the dielectric layer thickness, manifested by a red shift that increases with decreasing dielectric layer thickness. [29]

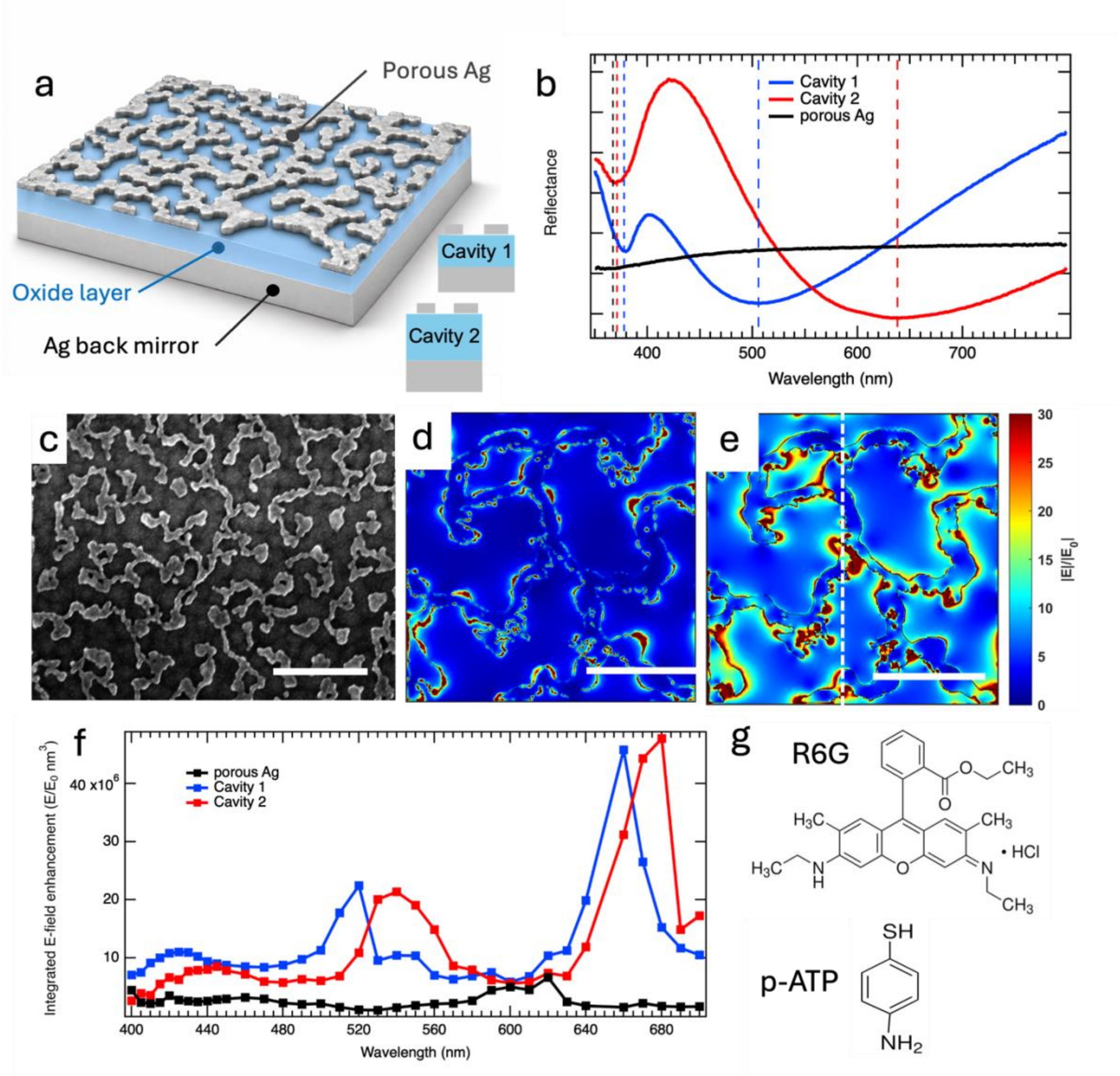

**Figure 1.** (a) Illustration of the nanoporous metal on dielectric metal (NPMoDM) structure, in which the low-frequency resonance can be tuned via the thickness of the dielectric layer (Cavity 1 and Cavity 2). (b) Reflectance spectra measured from a bare porous Ag film (black), and from NPMoDM structures with different alumina thickness (Cavity 1: 60 nm; Cavity 2: 80 nm). The dashed vertical lines indicate the position of the minima. c) SEM image of a typical nanoporous silver layer. Scale bar is 250 nm. (d,e) Simulated electric nearfield map of the p-Ag film at 400 nm, (d) and the NPMoDM structure (e) at the resonance wavelength of 512 nm (Cavity 1). Scale bars are 100 nm. The vertical dashed line indicates the cut taken for the cross sections in Figure 2. See **Figure S1** for data on Cavity 2. f) Wavelength dependence of the integrated near field intensity

($|E|/|E_0|$) over the volume of the porous Ag layer. g) Structure of the analyte molecules R6G and p-ATP.

**Figure 1c** shows a scanning electron microscopy (SEM) image of a nanoporous Ag film, evidencing a morphology of gaps in the nanoporous film on length scales of few tens of nanometers. The spatial near-field enhancement, calculated by finite elements methods (COMSOL) simulations, resulting from the nanoporous metal layer is shown in **Figure 1d**, and that of the NPMoDM structure (Cavity 1) in **Figure 1e**. Comparison of Figures 1d and 1e shows a strongly increased near-field enhancement of the NPMoDM structure with respect to the bare nanoporous Ag film (data for the Cavity 2 structure is shown in **Figure S1)**. Figure 1f plots the spectral dispersion of the nearfield enhancement (obtained by integrating over the volume of the porous metal layer), manifesting two peaks that are slightly red shifted for the thicker Cavity 2 with respect to Cavity 1. These simulations showcase the different behavior of the plasmonic hotspots with respect to the cavity resonances, indicating the hybrid character of the NPMoDM substrates.[30] We find that for Cavity 1 we are in resonance with both nearfield and far-field (reflectance) enhancement, while Cavity 2 is out of resonance with the nearfield, but in resonance with the far-field at 633 nm. We attribute the red-shift observed for the peaks in nearfield enhancement observed for Cavity 2 in Figure 1f to the different dielectric environment induced by the thicker underlying cavity. This spectral tunability of the field enhancement via the oxide layer thickness and the morphology of the plasmonic nanoporous film provides highly appealing opportunities to tailor the substrate response to match the available laser lines in the Raman spectroscopy investigations.

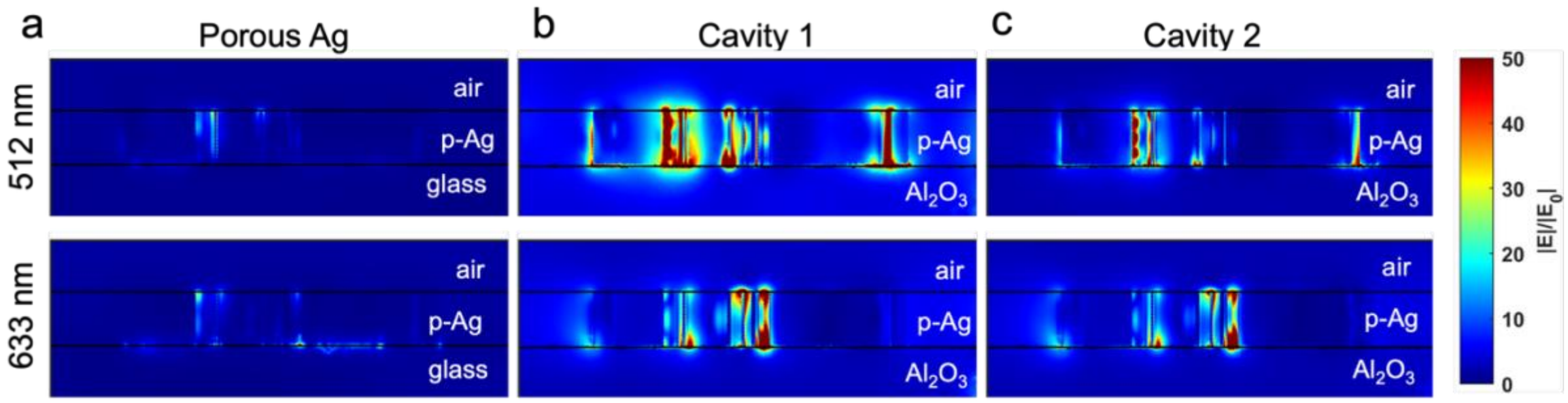

**Figure 2.** Cross sections of the simulated electric nearfield for (a) porous Ag, (b) Cavity 1, and (c) Cavity 2 along the cut indicated by the dashed white line in Figure 2e. Incident light excitation from the top at 512 nm (top row) and 633 nm (bottom row) wavelength.

**Figure 2** shows cross section maps of the calculated nearfield along the z-y directions for the bare porous Ag layer, Cavity 1, and Cavity 2 at the wavelengths available in our Raman setup (512 nm and 633 nm). We find that for the bare Ag layer the enhancement is mainly localised on the surface of the film, while the NPMoDM structure strongly enhances the penetration of the nearfield enhancement into the bulk porous film, leading to a 3D distribution of hot spots. Such increased near-field enhancement intensity and volume overlaps extremely well with the binding sites of the analytes, and therefore is highly favorable for ultrasensitive SERS.

For the SERS performance study, we use R6G (resonant with green light excitation) and p-ATP (non-resonant in the visible) as representative analytes to functionalize our substrates. Details of the analyte functionalisation are reported in the Experimental Section. We first fix the concentration of the analytes to 1 µM to directly compare the signal strength of the different substrates under excitation with green (514 nm) and red (633 nm) laser sources. In **Figure 3** we plot all panels on the same scale for a direct visual comparison of the Raman signal strength under the various conditions. As expected, we obtain the strong enhancement when the substrate cavity is tuned to the excitation laser wavelength (Cavity 1 in **Figure 3a,c** and Cavity 2 in **Figure 3b,d**).

We observe the strongest enhancement for the R6G with the 514 nm laser excitation and Cavity 1 that matches the resonance condition (Figure 3a), and for p-ATP the strongest Raman signal under 633 nm excitation with Cavity 2 (Figure 3d). In all cases we find a strongly enhanced Raman signal compared to the single nanoporous silver (p-Ag) film. By taking the signal from the nanoporous Ag film as reference, we can quantify an enhancement factor (EF= $I_{Cavity}/I_{p\text{-}Ag}$)) evaluated at prominent Raman peaks of the analyte as specified in the panels of Figure 3, however, we note that this EF strongly depends on the performance of the p-Ag substrate with the 514 nm and 633 nm excitation wavelengths.

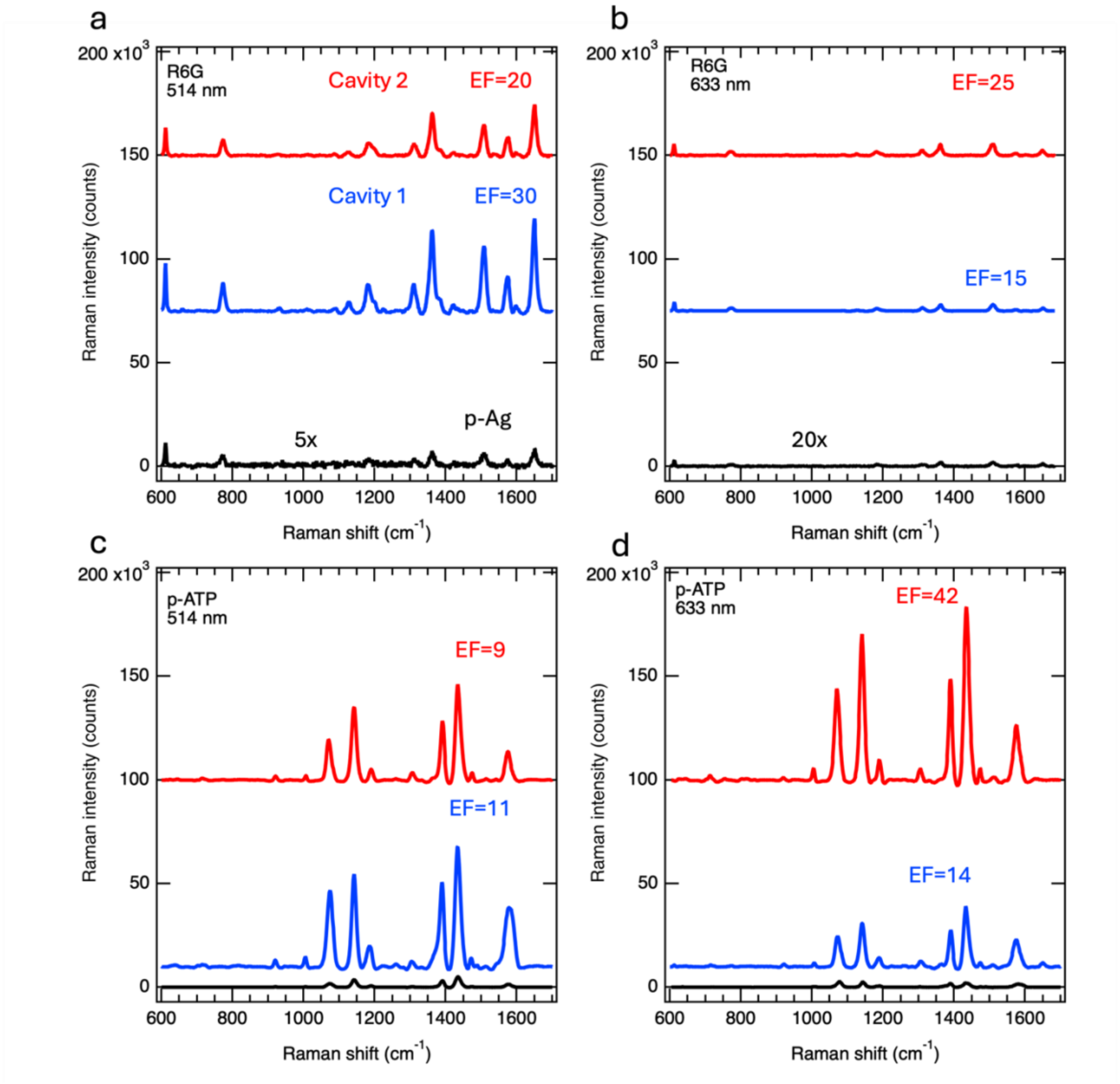

**Figure 3.** SERS spectra of R6G and p-ATP functionalized with 1 µM dilution on P-Ag reference substrates and P-MDM cavities measured with different laser excitation wavelength of 514 nm and 633 nm. **(a-b)** Rhodamine 6G (R6G) and **(c-d)** p-amino thiophenol (p-ATP). Laser power was 0.63 mW for 514 nm and 2.08 mW for 633 nm, and signal integration time was 1s for R6G and 10s for p-ATP. The spectra were recorded with a 5x objective. The signal of p-Ag in (a) and (b) is multiplied by the indicated factors, and spectra are vertically offset for clarity. Color coding of blue for Cavity 1 and red for Cavity 2 holds for all panels.

Next, we elaborate on the sensitivity limit of the NPMoMD substrates by SERS measurements using different analyte solution concentrations, for which we reduced the molar concentration in the substrate functionalisation process from 1 nanomolar ($10^{-9}$M) in steps of 1000 down to 1 zeptomolar ($10^{-21}$M). A rough estimation of the molecule density on the substrate (see SI for details) shows that at aM and zM concentrations we can expect to find only one or few molecules in the Raman spot. Accordingly, for aM concentration roughly only one out of ten positions produced a meaningful Raman signal of the analyte, while for zM solutions we overall only found a few regions on the sample surface that yielded signal of the analyte.

**Figure 4a** demonstrates that we can resolve the Raman signal of R6G under resonant conditions down to the above specified single molecule range (aM and zM concentrations) of the analyte. The most prominent peak at 1650 $cm^{-1}$ corresponds to the xanthene ring stretch. Other characteristic peaks are present at 610 $cm^{-1}$ (in-plane xanthene rings), 770 $cm^{-1}$ (C-H out-of-plane bend), 1180 $cm^{-1}$ (C-H in-plane bend), 1310 $cm^{-1}$(C-O-C stretch), 1360 $cm^{-1}$ (aromatic C-C stretching). Comparing spectra at different spot locations recorded from samples functionalized with zM concentration, we observe fluctuations in the peak occurrence and peak amplitude and width, while the characteristic molecular fingerprint is still evident, which is a typical signature of single molecule detection. Such spectral differences arise from the different orientations of the R6G molecules on the Ag nanostructures[31] in the diverse probing spots. [32, 33] We conclude that with our NPMoDM substrates we can achieve single molecule detection sensitivity under resonant conditions for R6G as analyte (see **Figure S2** for data recorded with Cavity 2 under 633 nm excitation). **Figure S3** reports spectra recorded at aM concentrations from different Raman spot locations on the substrate surface.

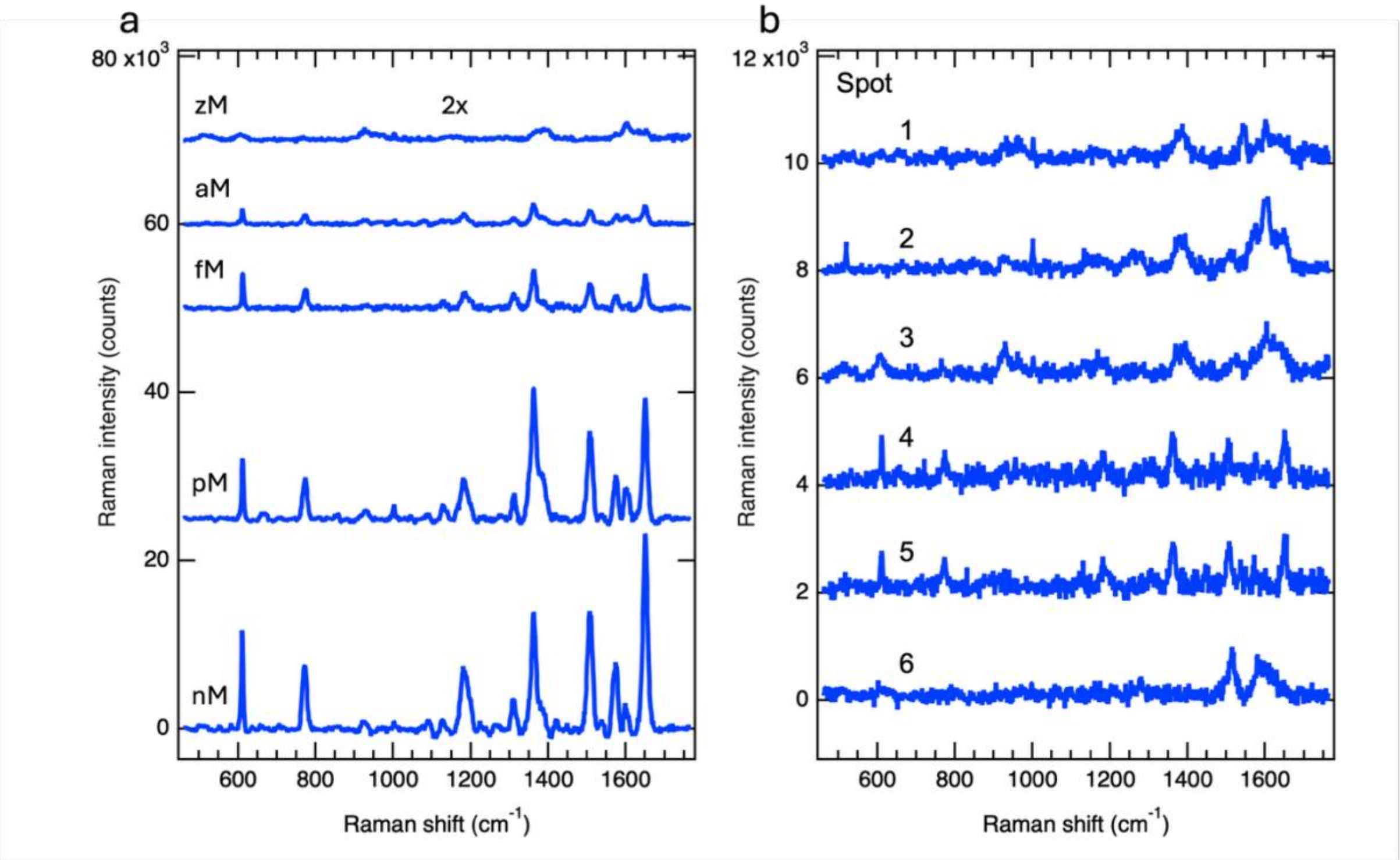

**Figure 4. (a)** SERS spectra recorded with Cavity 1 for different R6G analyte dilutions. The excitation wavelength of 514 nm was in resonance with the cavity (see Figure 1). **(b)** SERS spectra recorded from six different spot positions on the substrate surface with 1 zM analyte concentration in the functionalisation. Laser power was 0.63 mW, signal integration time was 10s, and spectra were acquired with 5x objective. Spectra are vertically offset for clarity. Data recorded with Cavity 2 from R6G at different dilutions is reported in **Figure S2** and shows that single molecule detection is not reliably achieved with R6G on Cavity 2.

We proceed to test the SERS functionality of our substrates for non-resonant analytes, using p-ATP. [34-36] Here we follow a similar functionalisation procedure as for R6G but use ethanol instead of water, spanning analyte dilutions from nM to zM concentration. We note that we can expect conformationally stable chemisorption through sulfur bonding with p-ATP on silver.[37] **Figure 5** reports the SERS data recorded at different concentrations for both NPMoDM cavities that were illuminated with the laser lines that correspond to their spectral resonances (that is 514 nm for Cavity 1 and 633 nm for Cavity 2). We are able to discern the Raman signatures of p-ATP down to aM and zM analyte concentration for both cavities, with characteristic peaks at 1004 $cm^{-1}$ (ring breathing mode), 1070 $cm^{-1}$ (C-S stretching), 1190 $cm^{-1}$ (in-plane C-H deformation), and 1575 $cm^{-1}$ (C-C stretching, benzene ring). We note that the bands at 1140 $cm^{-1}$, 1390 $cm^{-1}$, 1435 $cm^{-1}$ are representative of possible transformations of p-ATP to dimercaptoazobenzene (DMAB),[38] however, this is not the subject of our investigation and it does not affect the evidence on the ultrahigh sensitivity of our SERS substrates. Probing multiple spots at zM concentration yields relatively reproducible spectra for Cavity 2 (**Figure 5d**), while with Cavity 1 notable fluctuations occur (**Figure 5c**). This behavior could be related to laser illumination/hotspot induced effects on the molecules that are different for both cavities.

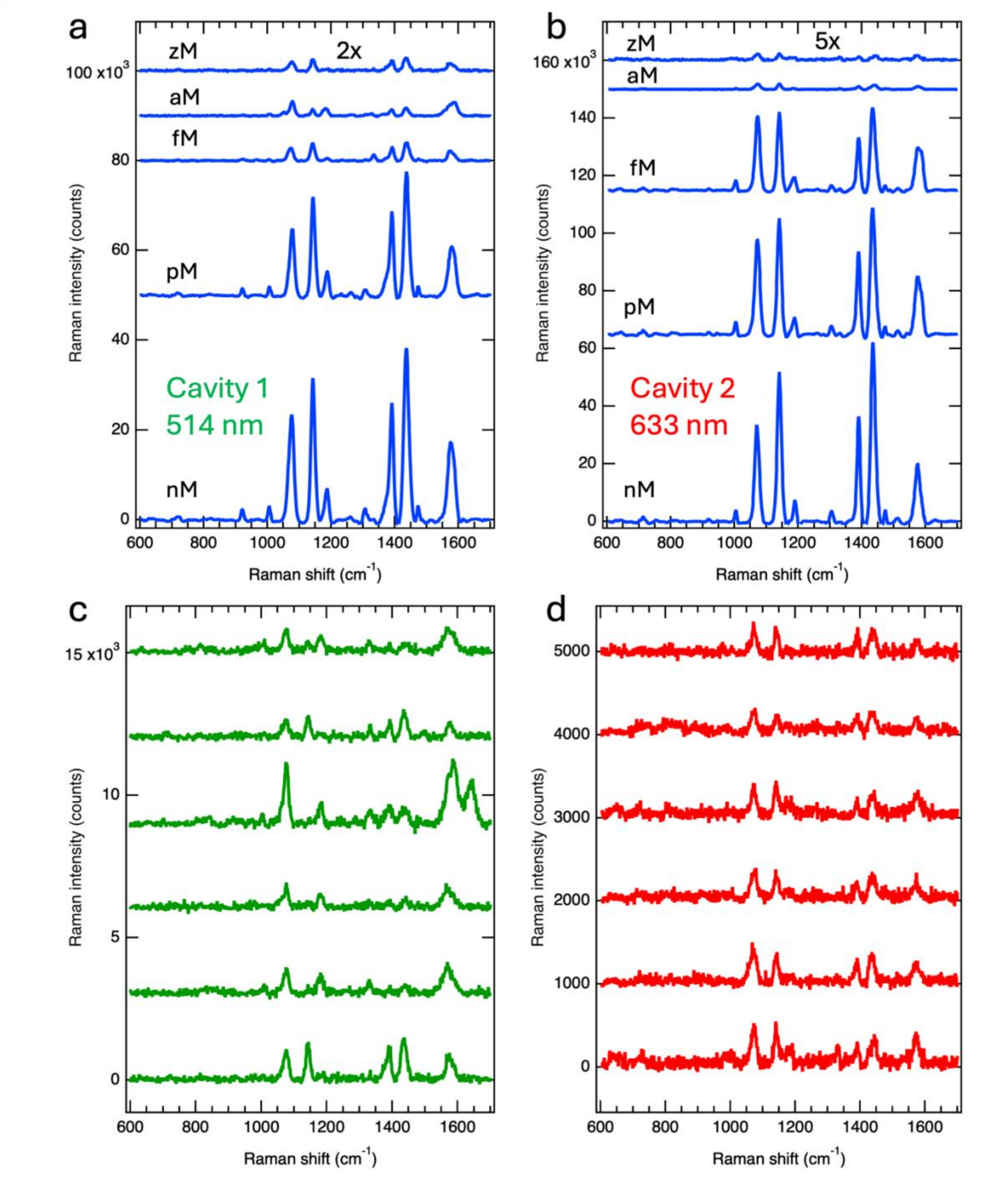

**Figure 5. (a)** SERS spectra of p-ATP recorded with Cavity 1 at 514 nm excitation and **(b)** with Cavity 2 at 633 nm excitation, **(c, d)** SERS spectra of Cavity 1 and 2 acquired from different spots at 1zM concentration under 514 nm and 633 nm laser excitation, respectively. Laser power was set to 0.63 mW for 514 nm and 2.08 mW for 633 nm, signal integration time was 10 s in all cases, and spectra were recorded with a 5x objective. Spectra are vertically offset for clarity.

Next, we investigate how the SERS signal scales with the molecular concentration that we used in the substrate functionalization. Figure 6 plots the Raman intensities obtained from a set of peaks extracted from the data in Figure 4, Figure S3 and Figure 5 versus the concentration shown on a logarithmic scale. Interestingly, we observe different trends for Cavity 1 and Cavity 2. For Cavity 1 the shape of the concentration-response curve agrees with a sigmoidal shape as it has been reported for substrates with a population of plasmonic hot spots. [39, 40] This is in agreement with the pronounced resonance and strong localized field enhancement of Cavity 1 that we report in Figure 1, and substrate coverage could follow a Langmuir-like behavior. Instead for Cavity 2 the signal increase is strongly sub-linear, which is in line with the weaker nearfield enhancement, as evidenced in Figure 2c.

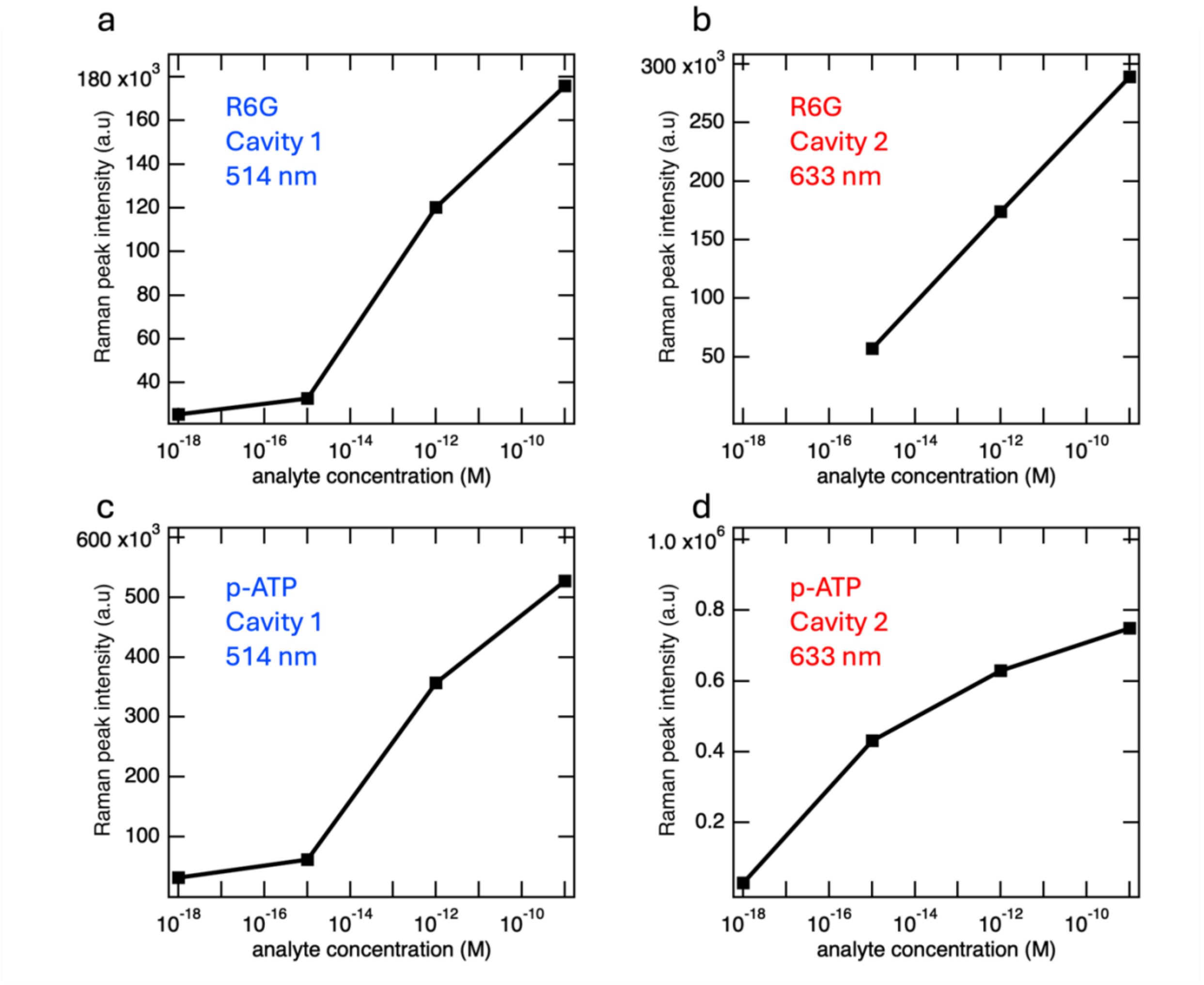

**Figure 6.** Raman signal intensity (peak area) versus the dilution concentration (in M) for the different analytes and cavity systems (data from spectra in Figure 3a, Figure S2a, and Figure 4a,b). The peak intensity was evaluated and then averaged from four dominant peaks in each spectrum (at 611, 772, 1310 and 1650 $cm^{-1}$ for R6G and at 1075, 1143, 1390, 1436 $cm^{-1}$ for p-ATP).

In conclusion, we demonstrated how highly sensitive and frequency-tunable SERS substrates can be obtained by combining porous Ag films with an underlying dielectric-metal bilayer that strongly enhances the electrical near-field, and results in hybrid plasmonic-cavity resonances. Single molecule detection capability of the substrates was demonstrated by detecting fluctuations in peak width and intensity that originate from different binding geometries of the molecule to the Ag nanostructures. Such ultrahigh sensitivity obtained with a simple, scalable and economic fabrication of the substrates will make them attractive for a large variety of SERS applications. Furthermore, the readily accessible and homogeneous surface facilitates analyte functionalization and Raman spectroscopy experiments. Biocompatibility and long-term stability of the substrates can be straightforwardly enhanced by using the Au/Ag bilayers in the nanoporous metal film that we recently reported. [25] Beyond SERS, such structures employing different metals can also be applied in photocatalytic applications. [41]

**Experimental Section.**

Substrate Preparation. Silicon substrates were cleaned by sonication in acetone and isopropanol for 10 minutes each followed by nitrogen blow dry. The multi-layer structure was deposited by electron beam evaporation. (1) First a bottom silver mirror layer with thickness of 100 nm was deposited followed by aluminium oxide layer ($Al_2O_3$). (2) Then polymer layer (PMMA) was spin coated on oxide layer and baked at 180 $^{o}$C. (3) Finally, semi-transparent top silver layer with a thickness of 13 nm was deposited with electron beam evaporator and the surface was etched using nitrogen plasma for 10 minutes at normal incidence. For the reference p-Ag substrate only steps (2) and (3) we performed.

Analyte Dilution. To prepare analytes, a 10 mM stock solution of Rhodamine 6G and p- Amino thiophenol in Milli-Q water and ethanol, respectively, was prepared with a total volume of 10 mL. A dilution series from 1 mM down to zeptomolar (zM) concentrations was then prepared through sequential 1:10 dilution steps. At each step, 1 mL of the previous solution was mixed into 10 mL of the corresponding solvent.

Surface functionalization and Raman measurements. Separate 8 mm × 8mm sized substrates were used for every concentration. Each substrate was immersed in 2 mL of the analyte solution for 20 minutes. After incubation, samples were swiftly blow-dried with nitrogen. The SERS spectra were recorded with a Renishaw InVia Raman spectrometer under ambient conditions with a 5x objective (N.A. 0.12) by using 514 nm (laser power at the sample position 0.63 mW) and 633 nm laser excitation (laser power at the sample position 2.08 mW), and diffracting grids with 2400 lines/mm and 1800 lines/mm, respectively. The measurements were acquired in back scattering geometry, with 1s exposure and different number of accumulations, as specified in the text.

Finite element methods modelling. To study the plasmonic properties of nanoporous silver film in combination with the underlying cavity, a numerical study of the electromagnetic response was conducted using the finite element method (FEM) with COMSOL Multiphysics. Following the procedure described in detail in our recently published works, [25, 42] the nanometer-sized pores and irregularities of nanoporous Ag were numerically reproduced using SEM images of the experimental samples. For this, the nanoporous Ag samples were processed through binarization and were extruded in the direction of the film thickness to reconstruct a realistic three-dimensional pore geometry reflecting the actual pore size distribution and nanoscale heterogeneities of the experimental structures. A representative periodic region measuring 250 × 250 $nm^2$ was modeled. The simulated models were illuminated at normal incidence by a plane wave with wavelengths of 514 and 633 nm, which corresponded to the excitation wavelengths used in the experiment. To assess the local field enhancement, the ratio of the norm of the electric field, $E/E_0$, in the presence of nanoporous metal to the amplitude of the incident field ($E_0$) was calculated.

ASSOCIATED CONTENT

**Supporting Information** with additional data from simulations and Raman spectroscopy experiments (PDF) is available free of charge.

AUTHOR INFORMATION

**Corresponding Author**

roman.krahne@iit.it

**Author Contributions**

The manuscript was written through contributions of all authors. All authors have given approval to the final version of the manuscript. ‡These authors contributed equally. (match statement to author names with a symbol)

ACKNOWLEDGMENT

We acknowledge funding by the European Union under Project 101131111 – DELIGHT, and HORIZON-Pathfinder-Open: DYNAMO, grant Agreement 101072818.

References.

TOC image

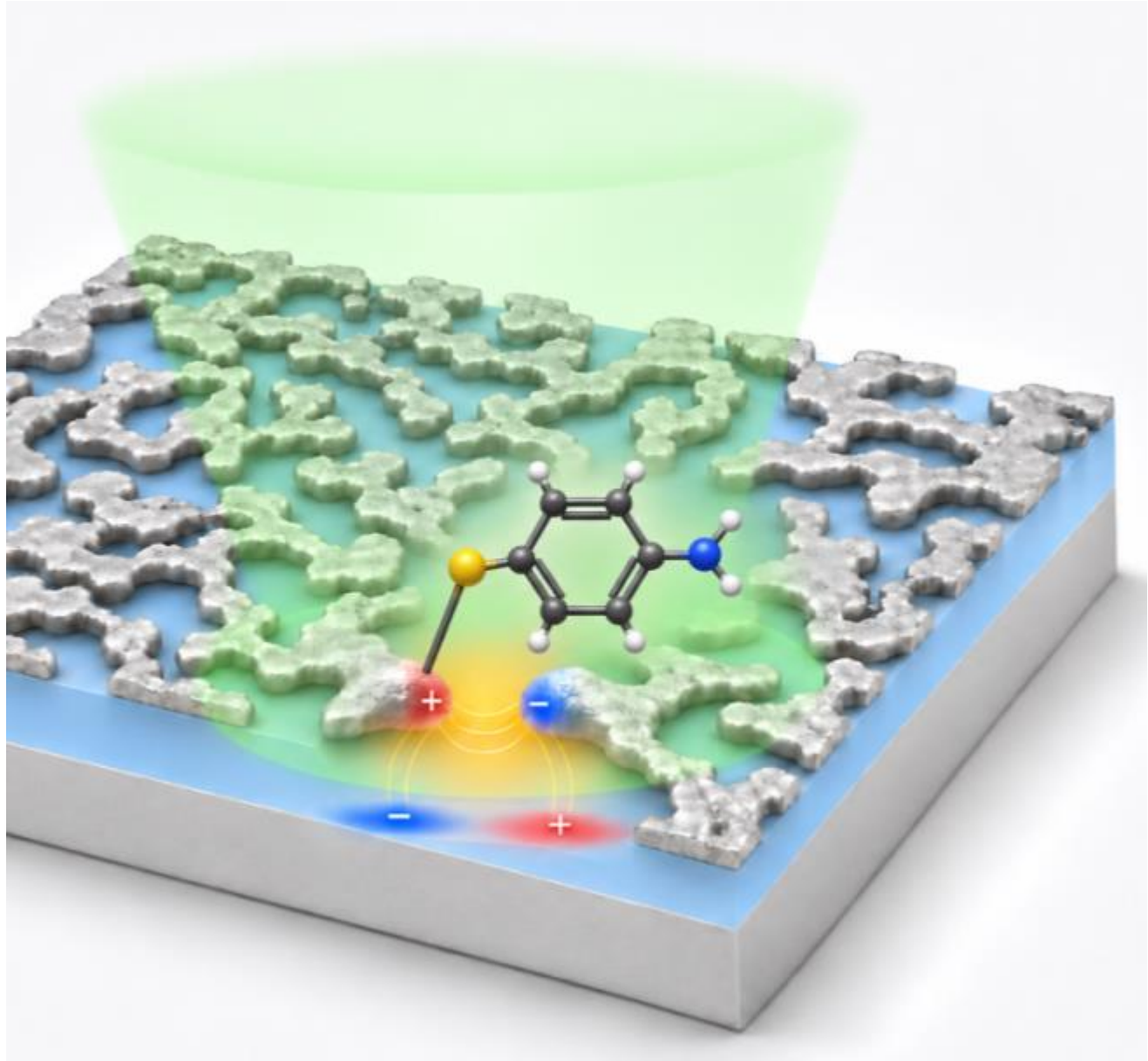

Supporting Information for

# Metal-Dielectric Bilayer Supported Nanoporous Silver Substrates for Ultrasensitive Surface-Enhanced Raman Scattering

*Maryam Naseer, Anastasiia Sapunova, Ali Douaki, Vincenzo Caligiuri, Silvia Dante, Denis Garoli, and Roman Krahne*

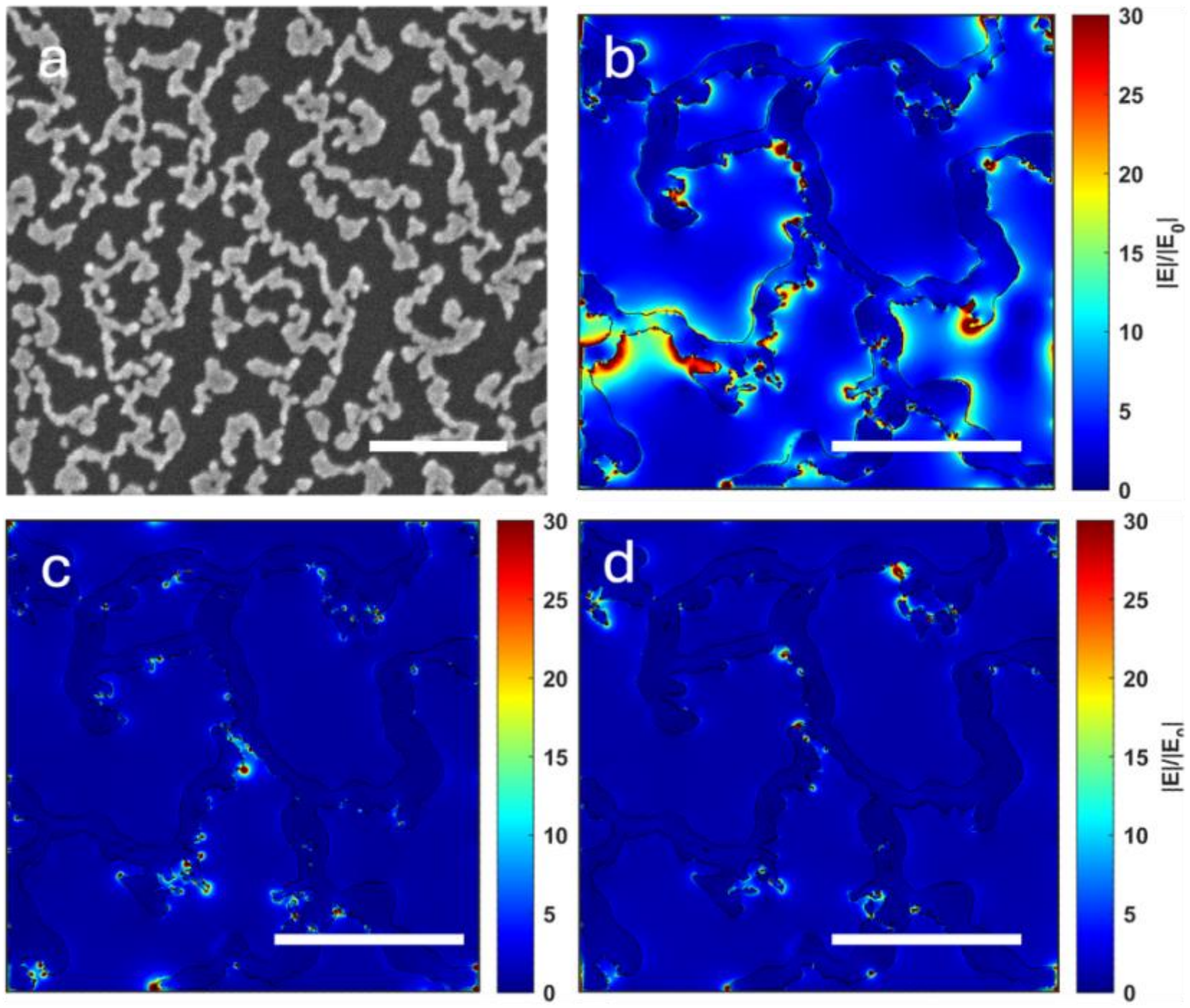

**Figure S1.** a) SEM image of the porous silver layer used to  extract the model for the simulations. Scale bar is 250 nm. (b-d) Simulated electric nearfield distribution of the NPMoDM structure (b) at the resonance wavelength of 633 nm (Cavity 2), and the porous Ag film at 512 nm (c) and 633 nm (d). Scale bars are 100 nm.

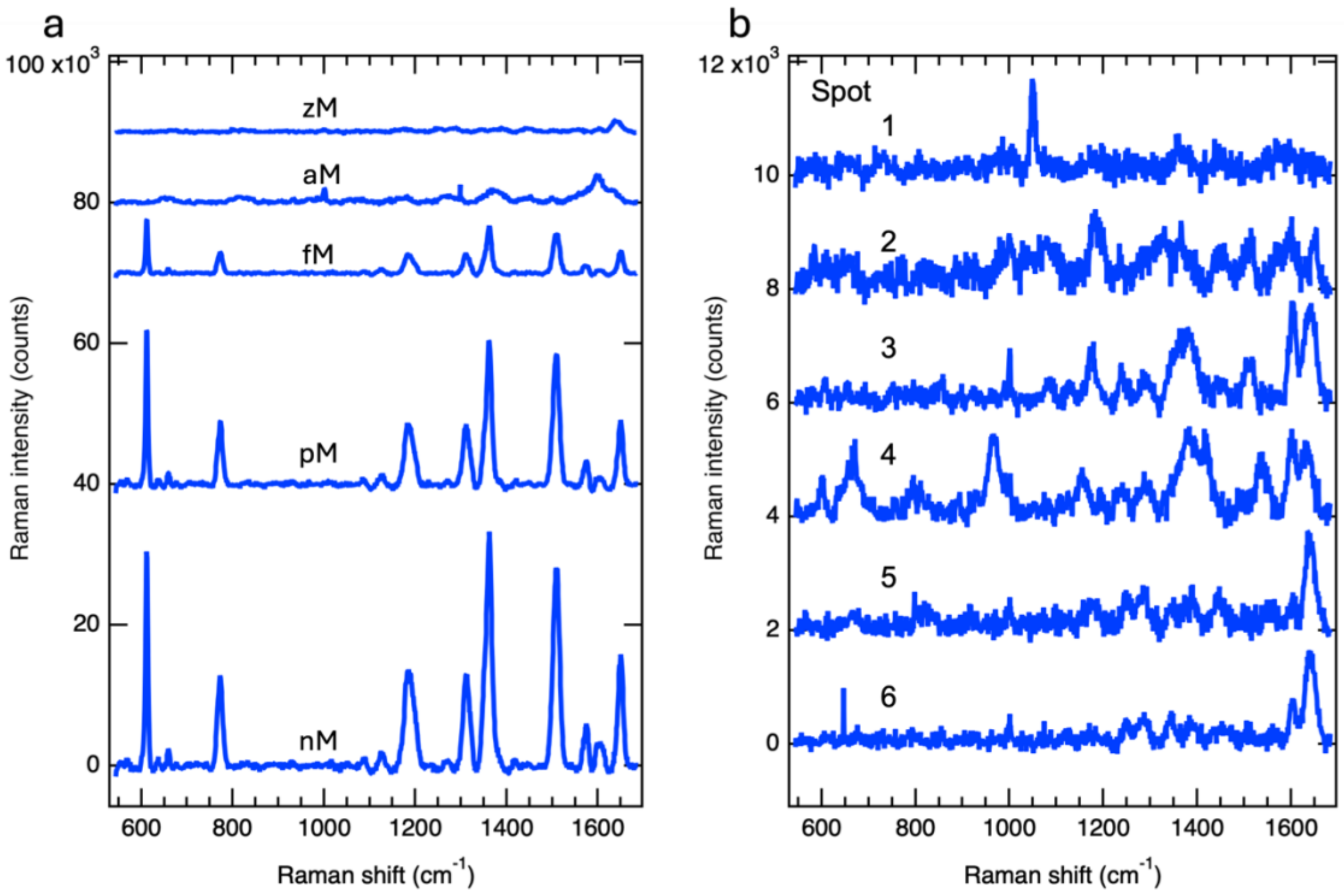

**Figure S2. (a)** SERS spectra recorded with Cavity 2 for different R6G analyte dilutions. The excitation wavelength of 633 nm was in resonance with the cavity (see Figure 1 in the main text). **(b)** SERS spectra recorded from 6 different spot positions on the substrate surface at 1zM concentration. Data was recorded with a 5x objective, laser power at the sample position was 0.63 mW, and signal integration time was 20s. We find that with aM and zM concentration the Raman signal is not always corresponding to the expected spectral shape of R6G, thus under these conditions the probing single/few molecules is not reliable. Spectra are vertically offset for clarity.

Estimation of molecule density on the substrate surface. Since we do not have direct access to the number of molecules that are active in our Raman experiments, we make the following estimate of the average number of molecules per $mm^2$ on the substrate that uses strongly overestimated numbers for the relevant solution volume. For this we approximate the number of the molecules that potentially can reach the sample surface by considering a diffusion length of 1 mm for R6G molecules (diffusion coefficient in water at 25°C is 414 $\mu m^2/s$) [1] and 2 mm of p-ATP molecules (diffusion coefficient in ethanol at 25°C estimated as 1000 $\mu m^2/s$). [2] This corresponds to the full 20 min immersion time in the incubation process. The height of the solution in the used glass beaker was 3 mm and the substrate size was 8 mm x 8mm. For a very conservative (upper estimate) we therefore consider the full height of the solution volume (3 mm) and a lateral size of the substrate extended by the diffusion length. This yields a volume of (8+1+1) mm * (8+1+1) mm * 3 mm = 300 $mm^3$ for the R6G molecules that can reach the substrate surface, and 12mm*12mm*3mm=432 $mm^3$ for p-ATP. With this overestimated volume we potentially compensate for the deposition of molecules during the immersion, extraction and blow-drying process. From the molar concentrations we can directly obtain the number of molecules in this volume, and assuming that all molecules are deposited uniformly on the surface, we can derive the molecule density per $mm^2$ on our substrates and in a spot of ca. 5-6 µm radius (20-30 $\mu m^2$) corresponding to the probed micro-Raman region, as reported in **Table 1**. We find that for aM and zM concentrations the average number of molecules in the probe spot is much less than unity, therefore these dilutions should fall into the single molecule detection range.

**Table 1.** Number of molecules per substrate area and in the probe spot of R6G and p-ATP for the different dilutions.

| **Analyte dilution concentration** | **R6G Molecules/mm²** | **R6G Molecules/Raman spot (20 µm²)** | **p-ATP Molecules/mm²** | **p-ATP Molecules/Raman spot (20 µm²)** |
|---|---|---|---|---|
| **µM** | $10^{12}$ | $2 \times 10^{7}$ | $1.5 \times 10^{12}$ | $3 \times 10^{7}$ |
| **nM** | $10^{9}$ | $2 \times 10^{4}$ | $1.5 \times 10^{9}$ | $3 \times 10^{4}$ |
| **fM** | $10^{6}$ | 20 | $1.5 \times 10^{6}$ | 30 |
| **aM** | $10^{3}$ | $2 \times 10^{-2}$ | $1.5 \times 10^{3}$ | $3 \times 10^{-2}$ |
| **zM** | 10 | $2 \times 10^{-5}$ | 15 | $3 \times 10^{-5}$ |

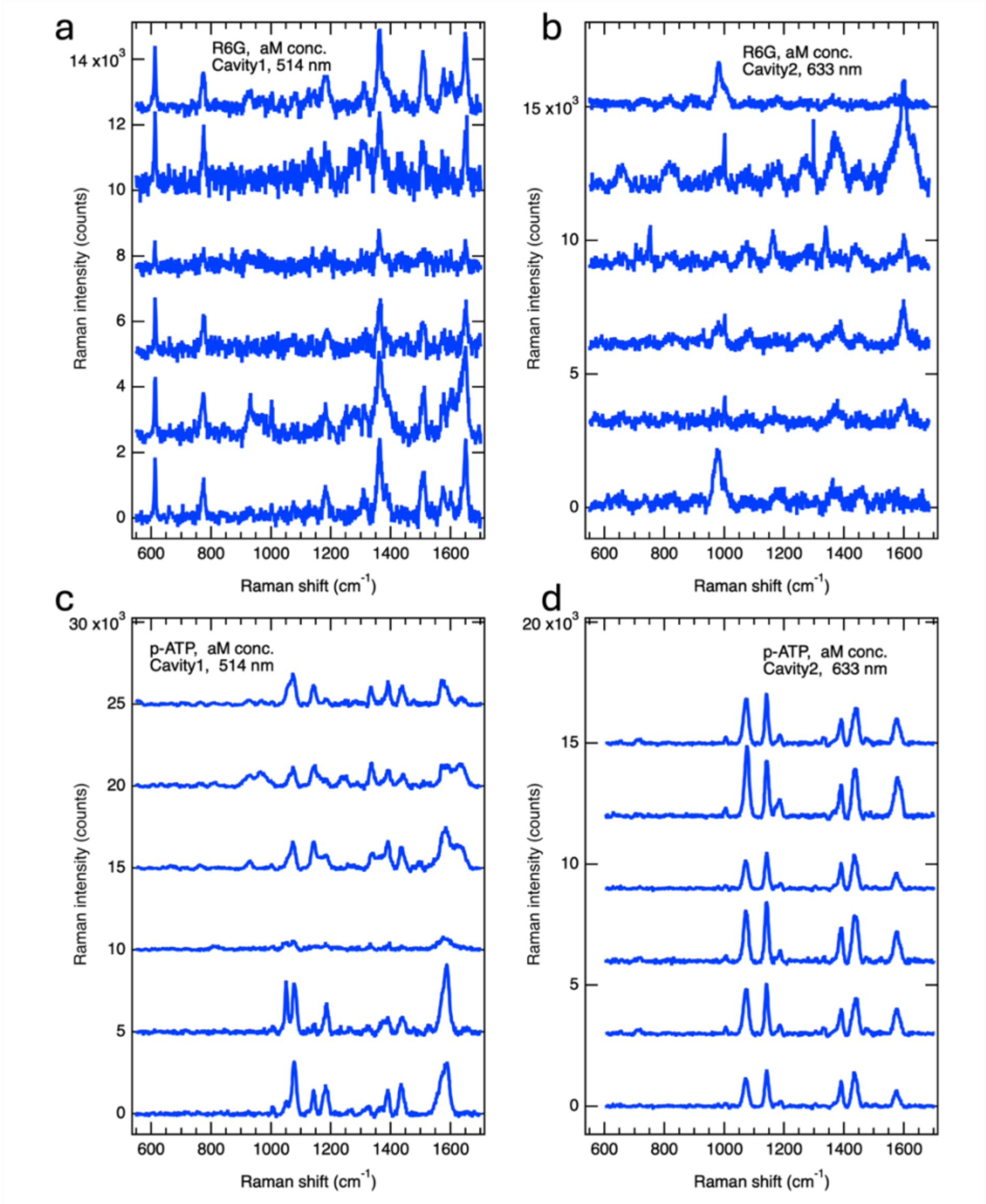

**Figure S3.** Raman spectra recorded from different locations of samples prepared with aM concentrations of the analytes. Data was recorded with a 5x objective, laser power was 2.08 mW, and integration time was 10 s.

**References.**